\documentclass[prl,showpacs,groupedaddress,twocolumn,longbibliography]{revtex4-2}
\usepackage{graphicx}
\usepackage{amsmath}
\usepackage{amssymb}
\usepackage{amstext}
\usepackage{graphicx}
\usepackage{mathtools}
\usepackage{bm}
\usepackage{hyperref}
\hypersetup{
	breaklinks=true,
	colorlinks=true,
	allcolors=blue,
}

\newcommand{\vechat}[1]{\hat{\bm{#1}}}
\newcommand{\spk}[1]{#1}

\begin{document}

\title{Superconductivity-enhanced magnetic field noise}
\author{Shane P. Kelly}
\author{Yaroslav Tserkovnyak}

\affiliation{Department of Physics and Astronomy and Mani L. Bhaumik Institute for Theoretical Physics, University of California, Los Angeles, CA 90095}

\begin{abstract}
    We consider the stray magnetic field noise outside a two-dimensional superconductor.
    Our considerations are motivated by recent experiments, which observed an enhancement in the magnetic field noise below the superconducting critical temperature, based on the relaxation of diamond nitrogen-vacancy centers.
    Such enhancement is not captured by the standard two-fluid model for the superconducting state, recently proposed to explain such NV relaxometry experiments. 
    Instead, we show that a microscopic BCS theory captures such an enhancement, and we compare with a similar theory and phenomenon, known as the Hebel-Slichter peak (or coherence peak), observed in the relaxation of nuclear spins in the material.
    The primary difference is that the NV probes long-wavelength magnetic noise outside the sample, while the nuclear spin probes local hyperfine noise inside the sample. 
    Accordingly, the noise probed by the NV depends on its height and can probe, in pristine samples, the superfluid coherence length.
    Finally, we discuss potential avenues for NVs to probe unconventional superconductivity via deviations from the above BCS theory.
\end{abstract}

\maketitle

Recent progress in optics and nanofabrication have allowed the development of atomic qubits embedded in solids.
Color centers, such as Nitrogen-vacancy impurities in diamond~\cite{Doherty_2013}, vacancy centers in silicon carbide~\cite{Nagy:2019aa}, or defects in hexagonal boron nitride~\cite{Gottscholl:2020aa,Huang:2022aa,Lee:2022aa,Zhou:aa} can behave as optically addressable qubits with long coherence times~\cite{Chatterjee:2021aa}.
While such qubits might play a role in future advanced quantum technologies such as quantum computers~\cite{Ladd:2010aa}, or quantum signal relays~\cite{Kimble_2008}, currently they are being developed as precision magnetic-field sensors~\cite{RevModPhys.89.035002,Taylor:2008aa,Maze:2008aa,kolkowitzProbingJohnsonNoise2015}.
In a static mode~\cite{Maze:2008aa,Taylor:2008aa,Doherty_2013}, shifts in magnetic levels of the impurities probe static magnetic fields and can be used to sense current and magnetization textures~\cite{Tetienne:2014aa,Li:2023aa,Palm:2024aa,Tetienne:aa}, aiding in the development of spintronic devices~\cite{Lee-Wong:2020aa,10.1063/5.0150709,McLaughlin:2023aa}.
While in a dynamical mode~\cite{PhysRevLett.108.197601,Langsjoen_2012,Rovny:2022aa}, relaxation of the qubit state probes magnetic field noise which can be used to infer microwave conductivities~\cite{kolkowitzProbingJohnsonNoise2015,Ariyaratne:2018aa} and probe magnon dynamics in magnetic materials~\cite{Sar:2015aa,Du:2017aa,Purser:2020aa,Wang:aa,McLaughlin:2022aa}.

Recent theoretical results have also suggested NV sensing of stray magnetic field noise may be a fruitful probe of unconventional superconductivity~\cite{Dolgirev_2022,curtis2024probingberezinskiikosterlitzthoulessvortexunbinding}.
These analyses consider a standard two-fluid model of normal and superconducting electrons in which the relative densities of the fluids are determined by free energy considerations.
In this model the normal-fluid contributes to noise while the superconducting fluid does not.
When the temperature decreases below criticality, the normal-fluid density is exponentially suppressed in temperature leading to an exponential suppression of noise.
In contrast, a recent experiment~\cite{experiment} found an enhancement in magnetic field noise as the temperature is reduced below the critical temperature.

In this work, we show that this enhancement of the $T_1^{-1}$ relaxation rate can be captured by the BCS model of superconductivity.
The BCS model of superconductivity involves the condensation of Cooper pairs and the associated emergence of a superconducting band gap at the Fermi surface.
In a pristine $s$-wave superconductor, this restructured quasiparticle dispersion results in a singular density of quasiparticle states.
For real materials, the divergence is cured by impurities, but the impact of the enhanced density of states can have significant consequences.
In particular, local spin noise increases below criticality, and leads to the enhanced nuclear spin relaxation rates that have been observed in early experiments~\cite{hebelNuclearSpinRelaxation1959}.

It is therefore natural that a similar enhancement may occur in the stray magnetic field noise outside a material.
Nonetheless, such an enhancement is not guaranteed.
For example, charge density noise is not enhanced.
Furthermore, the magnetic noise probed by the NV depends on its height above the material: it maybe the case that noise is enhanced but only very close to the material.
We therefore present in this work the stray magnetic field noise predicted by a microscopic BCS model of $s$-wave superconductivity.

We find that, similar to nuclear relaxation rates, the BCS model predicts an enhancement in the $T_1^{-1}$ relaxation rate for the NV below the superconducting critical temperature, and at reasonable heights above the material.
The main difference, is that nuclear relaxation is due to local magnetic field noise inside the sample, while the relaxation for the NV is due to long wavelength noise outside the sample.
Outside the material, magnetic field noise is sourced by current fluctuations, while inside the material, nuclear spins probe electron-spin noise via the hyperfine interaction.
We also determine the height dependence of the magnetic field noise.  
Close to the material, the magnetic field noise is enhanced by a factor similar to the factor for nuclear spins.
While for distances longer than the superconducting coherence length, the enhancement factor is logarithmically decreasing.

\begin{figure}[ht]
    \centering
    \includegraphics[width=1\columnwidth]{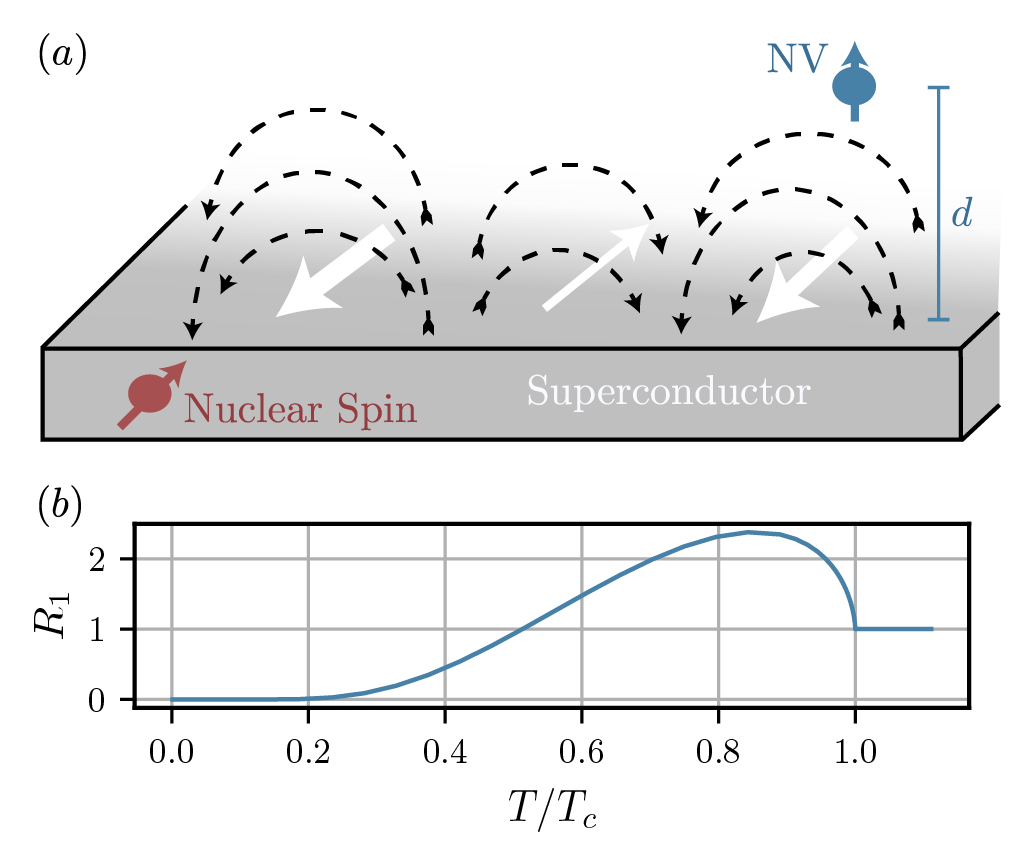}
    \caption{
        (a) Cartoon of magnetic noise sensed by a diamond NV center placed above a superconducting material, and by a nuclear spin located inside the sample. Current fluctuations are the primary source of the magnetic field noise probed by the NV and are depicted in white.
        (b) Temperature dependence of the NV relaxation rate, $T_{1}^{-1}(T)$, as a function of temperature $T$ relative to the expectation for a normal-state Fermi gas at the same temperature.
        In the figure we plot the factor, $R_{1}(T)=T_{1}^{-1}(T)T_c/T_{1}^{-1}(T_c)T$, where $T_1^{-1}(T_c) T/T_c$ is the linear in temperature expectation for the normal state~(See Eq.~\eqref{eq:normal} \spk{and Eq.~\eqref{eq:R_1})}. 
        Just below the critical temperature, magnetic noise sensed by the NV relaxation rate is enhanced, $R_1>1$.
        \spk{In this figure, the critical temperature is approximately $T_c\approx 10K$, the NV frequency is $\hbar\omega/k_B\approx 0.1 K$, and the NV height is less than the coherence length.}
    }
    \label{fig:summary}
\end{figure}

\textit{Coherence peak and singular density of states.---}
In the inspirational experiment~\cite{experiment}, an NV is placed at a fixed height, $d$, above a superconducting material~(see Fig.~\ref{fig:summary}(a)), and is prepared in an excited state with a magnetic-dipole moment.
The relaxation rate $T_1^{-1}$ was measured, and a temperature dependence similar to that in Fig.~\ref{fig:summary}(b) was observed.
This observation indicates an enhancement in magnetic field noise just below the critical temperature: the $T_1^{-1}$ relaxation rate and magnetic field noise are related via a Markovian approximation for the NV dynamics~\cite{flebusQuantumImpurityRelaxometryMagnetization2018,PhysRevB.98.195433,chatterjeeSinglespinQubitMagnetic2022,Dolgirev_2022,fangGeneralizedModelMagnon2022,zhangFlavorsMagneticNoise2022,PhysRevB.106.115108,zouBellstateGenerationSpin2022,curtis2024probingberezinskiikosterlitzthoulessvortexunbinding,li2024solidstateplatformcooperativequantum,PhysRevB.95.155107}.
The rate from this approximation is found as
\begin{eqnarray}~\label{eq:decayrate}
    T_1^{-1}=  \left(\frac{g_{\text{nv}}\mu_B}{2\hbar}\right)^2C_{\{B^+,B^-\}}(\omega,d)
\end{eqnarray}
where $g_{\text{nv}}\approx 2$ is the NV $g$ factor, $\mu_B=e\hbar/2m c$ is the Bohr magneton, and $C_{\{B^+,B^-\}}(\omega,d)=\int dt e^{i\omega t}\left<\left\{B^{-}(\bm{r}_{\text{nv}},t),B^{+}(\bm{r}_{\text{nv}},0)\right\}\right>$ is the power spectrum between the magnetic fields \spk{$B^{\pm}(\bm{r}_{\text{nv}},t)=B^{x}(\bm{r}_{\text{nv}},t)\pm i B^{y}(\bm{r}_{\text{nv}},t)$} with opposite circular polarization.
The circular polarization $\pm$ is defined with respect to the direction of the NV magnetic dipole, which in this letter we take to be $\hat{\bm{z}}$ direction, perpendicular to the superconducting plane.

The stray magnetic fields are produced by both spin and current fluctuations of the nearby material.
For simplicity, we assume the material is homogeneous and isotropic, and we relate the magnetic field $\bm{B}(\bm{q},d)=\int d\bm{\rho} e^{-i\bm{\rho}\cdot\bm{q}}\bm{B}(\bm{\rho}+d\hat{\bm{z}})$ to both the two-dimensional magnetization associated with the spin density, $\bm{M}(\bm{q})$, and to the two-dimensional current density, $\bm{J}(\bm{q})$.
Here, $\bm{\rho}$ and $\bm{q}$ are vectors lying in the superconducting plane.
In the supplemental material~\cite{supplement}, we provide exact expressions, but the important features are captured by the relations 
\begin{eqnarray}~\label{eq:kernels}
    \left|\bm{B}(\bm{q},d)\right|\propto qe^{-qd} \left|\bm{M}(\bm{q})\right| ,\,\,\,
    \left|\bm{B}(\bm{q},d)\right|\propto\frac{e^{-qd}}{c}\left|\bm{J}(\bm{q})\right|,
\end{eqnarray}
where the proportionality is a numerical factor of order 1 and depends on both the direction of the wave vector $\bm{q}$ and the direction of either the current or magnetization densities at wave vector $\bm{q}$.
For both sources, the magnetic fields with wavelength smaller than the height of the NV, $\left|\bm{q}\right|^{-1}<d$ are exponentially suppressed.
Thus, the NV senses long wavelength current and spin fluctuations.

To model the current and spin noise, we assume the superconducting state is described by a BCS theory with an isotropic gap $\Delta$.
We consider a superconductor in the clean limit with mean free path longer than the coherence length $l>\xi_0=2E_F/\pi\Delta k_F$.
Furthermore, we assume the height of the NV is smaller than the mean free path $d<l$, such that the NV is only sensitive to ballistic fluctuations of the quasiparticle-hole pairs.

The quasiparticles of the BCS theory are formed through the hybridization of electrons and holes.
Their dispersion relation is given by $E_{\bm{k}}=\sqrt{(\xi_{\bm{k}}-E_F)^2+\Delta^2}$, where $\xi_{\bm{k}}$ is the dispersion of the normal-state electrons.
Notably, the density of states associated with this dispersion exhibits a square root singularity:
\begin{eqnarray}
    D_{\text{SC}}(E)=D_F \frac{E}{\sqrt{E^2-\Delta^2}},
\end{eqnarray}
where $D_F$ is the normal-state density of states at the Fermi surface.

\begin{figure}[t]
    \centering
    \includegraphics[width=1\columnwidth]{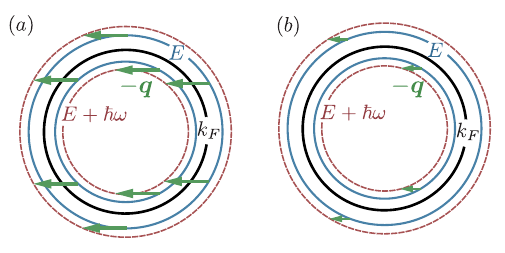}
    \caption{Cartoon depicting the solutions to the kinetic constraints $E=E_{\bm{k}_c}$ and $E+\hbar\omega=E_{\bm{k}_c-\bm{q}}$ for fixed energy $E$, frequency $\hbar\omega$ and wave vector $\bm{q}$. The green arrows show $-\bm{q}$ originating from the solutions $\bm{k}_c$. The black line shows the Fermi surface, while the solid blue line~(dashed red line) traces momenta at energy $E$~(at energy $E+\hbar\omega$). Figure~\ref{fig:kinetic_constraint}(a) shows the maximum eight solutions for a larger wave vector, while Fig.~\ref{fig:kinetic_constraint}(b) shows only four solutions for a smaller wave vector.}
    \label{fig:kinetic_constraint}
\end{figure}

Similar to previous results~\cite{Dolgirev_2022,zhangFlavorsMagneticNoise2022}, we find~\cite{supplement} that the magnetic field noise is dominated by transverse currents, $J_{\perp}(\bm{q})=\hat{\bm{z}}\cdot(\bm{J}\times\vechat{q})$.
In the ballistic limit, the transverse current noise, at frequency $\omega$ and wave vector $\bm{q}$, is due to fluctuating quasiparticle-hole pairs with energy difference $\hbar\omega$ and momentum difference $\hbar\bm{q}$.
An exact expression for these fluctuations constitutes our main result: 
\begin{eqnarray}~\label{eq:current_spectral_density_SC_main}
    C^{\text{SC}}_{\{J_{\perp},J_{\perp}\}}(\bm{q},\omega)= \int dt e^{i\omega t}\left<\{J_{\perp}(\bm{q},t)J_{\perp}(-\bm{q}))\}\right>=\\ \nonumber
    \int_\Delta^\infty dE D_{\text{SC}}(E) D_{SC}(E+\hbar\omega) g(E)\sum_{c}\frac{\hbar\left|\sin\theta_c\right|}{\left|\bm{k_c}\right|q}
\end{eqnarray}
which depends on a nonsingular function 
\begin{eqnarray*}
    g(E)=4\pi(ev_F)^2F_{+}(E,E+\hbar\omega)\!\!\!\!&[&\!\!\!\!n(E+\hbar\omega)\left(1-n(E)\right)  \\
     &+&\!\!\!\!\left(1-n(E+\hbar\omega)\right)n(E)]
\end{eqnarray*}
determined by the quasiparticle occupation via the Fermi-Dirac distribution $n(E)$. The function $g(E)$ also depends on the Fermi velocity $v_F$ and the coherence factor~\cite{tinkhamIntroductionSuperconductivity1996}, \spk{$F_{+}(E,E+\hbar\omega)=\left(u_{\bm{k}_c}u_{\bm{k}_c-\bm{q}}+v_{\bm{k}_c}v_{\bm{k}_c-\bm{q}} \right)^2$}, arising from the matrix elements relating quasiparticles to bare electrons and holes. For a \spk{circular Fermi surface, the quasiparticle weights $u_{\bm{k}}$ and $v_{\bm{k}}$ only depend on energy, and for relevant} energies, $E\approx \Delta$, the coherence factor is $F_{+}\approx 1$. Finally, $c$ indexes a sum over solutions $\bm{k}_c$ to the kinetic constraints,
\begin{eqnarray}~\label{eq:constraint}
    E=E_{\bm{k_c}},\,\,\, E+\hbar\omega=E_{\bm{k_c}-\bm{q}},
\end{eqnarray}
for a fixed energy $E$, frequency $\omega$ and wave vector $\bm{q}$. 
Example solutions to this equation are shown in Fig.~\ref{fig:kinetic_constraint}.
The angle $\theta_c$ is the angle between $\bm{q}$ and $\bm{k}_c$.

Equation~\eqref{eq:current_spectral_density_SC_main} thus takes contribution from all quasiparticle-hole pairs, with energy difference $\hbar\omega$ and momentum difference $\hbar\bm{q}$, that can be excited from the thermal cloud of quasiparticles.
At low temperature, the density of quasiparticles in the thermal cloud is exponentially suppressed resulting in an exponential suppression of noise.
This suppression is similar to the expectations from the two-fluid model and is captured in Eq.~\eqref{eq:current_spectral_density_SC_main} by the function $g(E)$: This function scales as $e^{-\Delta/k_BT}$ at low temperatures due to the Fermi-Dirac distributions.

At intermediate temperatures $T\lesssim T_c$, the thermal cloud has a density similar to the normal state, yielding $g(E)\approx (e v_F)^2$.
In this regime, the noise is now sensitive to the diverging superconducting density of states with a contribution $D_{SC}(E) D_{SC}(E+\hbar\omega)\approx 1/(E^2-\Delta^2)$.
This approximation is valid for $E>\Delta+\hbar\omega$ such that the integral yields a logarithmic singularity, $\ln(\Delta/\hbar \omega)$, for vanishing $\hbar\omega/\Delta$.

Accordingly, the phenomenon observed in the experiment, and shown in Fig.~\ref{fig:summary}(b), can be explained as follows.
Just below $T_c$, the gap is comparable to the NV frequency $\Delta\approx \hbar\omega$, and the current fluctuations are similar to the normal state.
As the temperature is reduced further, the gap increases and the logarithmic singularity becomes pronounced, enhancing fluctuations relative to the normal state.
Eventually, the gap saturates close to its zero temperature value.
Decreasing the temperature further only suppresses the quasiparticle cloud and the relaxation rate is suppressed exponentially with temperature.

\begin{figure}[t]
    \centering
    \includegraphics[width=1\columnwidth]{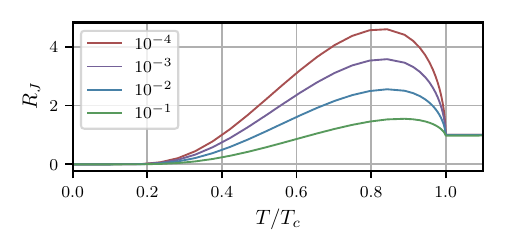}
    \caption{Current noise predicted by the BCS model relative to a normal-state Fermi gas at the same \spk{temperature: $R_{J}$ is defined in Eq.~\eqref{eq:R_J}. 
        Different curves correspond to different frequencies, where $\hbar\omega/k_B T$ is given in the legend.}
        The figure shows the noise at a wavelength smaller than the coherence length $1/q=10 \text{nm} <\xi_0$, but larger than the Fermi wavelength $2\pi/k_F\approx 0.2$nm. 
    The gap used was found from the self-consistent gap equation from BCS theory with $T_c=9$K. 
    Here,  the Fermi energy is chosen to be $E_F/k_B=20\times10^{4}$K such that the zero-temperature coherence length is $\xi_0\approx 500$nm.}
    \label{fig:q_dependence}
\end{figure}

In comparison, the normal-state current noise is decreasing linearly with temperature, and for a two-dimensional electron gas has the following form:
\begin{eqnarray}~\label{eq:normal}
    C^N_{\{J_{\perp},J_{\perp}\}}(\bm{q},\omega)=4(ev_F)^2n_e\frac{k_BT}{E_F}\frac{1}{v_F q}
\end{eqnarray}
where $n_e$ is the two-dimensional electron density, $q=\left|\bm{q}\right|$, and we have assumed that $q\ll k_F$ and $q\gg v_F/\omega$.
For $q$ outside this range, the current noise vanishes due to the kinetic constraint on quasiparticle-hole pairs in Eq.~\eqref{eq:constraint}.
This assumption is justified for typical NV experiments, which probe wave numbers of size $\left|\bm{q}\right|\approx 1/d$, and for which the NV height ranges from $d=50$nm to $d=10\mu m$, while $k_F^{-1}\approx 0.1$nm and $v_F/\omega\approx 1mm$.

In Fig.~\ref{fig:q_dependence}, we plot the ratio of the current noise in the superconductor relative to the noise for a normal-state Fermi gas with zero superconducting gap
\begin{equation}~\label{eq:R_J}
 \spk{R_{J}(q)=\frac{C^{SC}_{\{J_{\perp},J_{\perp}\}}(\bm{q},\omega)}{C^N_{\{J_{\perp},J_{\perp}\}}(\bm{q},\omega)}.}
\end{equation}
In the figure, the noise in the superconducting state $C^{SC}_{\{J_{\perp},J_{\perp}\}}(\bm{q},\omega)$ is evaluated using Eq.~\eqref{eq:current_spectral_density_SC_main} with the two-dimensional density of states $D_F = m/2\pi\hbar^2$.
Similarly, the noise for the normal state $C^N_{\{J_{\perp},J_{\perp}\}}(\bm{q},\omega)$ is evaluated using Eq.~\eqref{eq:normal}.
It is important to note that, Eq.~\eqref{eq:current_spectral_density_SC_main}, is exact while Eq.~\eqref{eq:normal} is an approximation from using $k_F \gg q\gg v_F/\omega$.
Fig.~\ref{fig:q_dependence} shows the enhancement in current noise as a function of temperature and frequency.
It confirms both the overall scaling in $\ln(\Delta/\hbar\omega)$, and the anticipated temperature dependence.

\textit{Local spin noise.---}
A similar picture holds for the spin noise probed by nuclear spins.
The main difference is that the nuclear spin probes local hyperfine noise that takes equal contributions from all momenta.
For concreteness consider the fluctuations of the spin density in the $\hat{\bm{z}}$ direction
\begin{eqnarray}\label{eq:spinnoise}
    C^{SC}_{\{M_z,M_z\}}(\omega)= \int dt e^{i\omega t}\left<\{M_z(\bm{r},t)M_{z}(\bm{r})\}\right>=\\ \nonumber
    \frac{2\mu_B^2}{(e v_F)^2}\int_\Delta^\infty dE D_{SC}(E) D_{SC}(E+\hbar\omega)g(E).
\end{eqnarray}
In comparison to the transverse current fluctuations in Eq.~\eqref{eq:current_spectral_density_SC_main}, the local spin fluctuations do not require the quasiparticle-hole pairs to be separated by a fixed momentum $\hbar\bm{q}$.
Thus, the momenta of the contributing quasiparticle-hole pairs $\bm{k_c}$ and $\bm{k_c}-\bm{q}$ do not affect the local spin fluctuations.
Nonetheless, the integral has similar dependence on the density of states and the quasiparticle occupation, resulting in similar physics.

\textit{Height dependence in pure crystals.---}
In contrast, the NV probes long wavelength noise, such that the momentum dependence of the kinetic constraint is relevant.
Concretely, the relaxation rate of the NV is given by
\begin{eqnarray}\label{eq:T1}
    T_1^{-1}&=&\left(\frac{g_{\text{nv}}\mu_B}{2\hbar}\right)^2\int d^2\bm{q}e^{-2dq}\frac{1}{c^2}C_{\{J_\perp,J_{\perp}\}}(\bm{q},\omega),
\end{eqnarray}
where the exponential factor $e^{-2dq}$ suppresses noise from short wavelength current fluctuations with wave number $\left|\bm{q}\right|>1/2d$.

For an NV at a height smaller than the superconducting coherence length, $d<\xi_0= 2E_F/\pi\Delta k_F$, the relaxation rate is enhanced in a fashion similar to of the nuclear spin~\footnote{Here we use the coherence length at zero temperature: The coherence peak occurs when $\Delta$ is comparable to its maximum, and thus the relevant finite temperature coherence length is already at its mininum $\xi_T\approx \xi_0$.}. 
In this limit, the NV probes current fluctuations at wave numbers in the range $k_f\gg\left|\bm{q}\right|>k_F \Delta/E_F$, and for which the factor $\hbar\sin\theta_c/k_cq$ is a constant function of energy. Thus, the transverse current noise in Eq.~\eqref{eq:current_spectral_density_SC_main} is proportional to the local spin noise in Eq.~\eqref{eq:spinnoise}, and the relative enhancement of the NV relaxation rate is similar to the enhancement for the nuclear spin relaxation rate.

For an NV further from the sample $d>\xi_0$, the NV relaxation rate decreases with increasing height.
The suppression is because the phase space satisfying the kinetic constraint shrinks, see Fig.~\ref{fig:kinetic_constraint}.
As we argue in the supplemental material~\cite{supplement}, the suppression is captured by a factor decreasing logarithmically with height.
\spk{Eventually, the kinetic constraint prevents quasiparticle pairs from forming at the band gap, near the singular density of states, and the noise in the superconducting state becomes suppressed as in the two-fluid model.}

\begin{figure}[t]
    \centering
    \includegraphics[width=1\columnwidth]{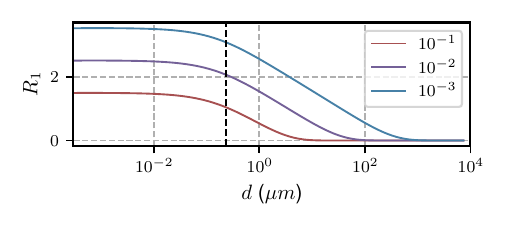}
    \caption{Height dependence of the NV relaxation rate predicted by the BCS theory relative to the rate for a normal-state Fermi gas at similar temperature: \spk{$R_1$ is defined~Eq.~\eqref{eq:R_1}}.
    The temperature is fixed at $T/T_c=0.9$, while the different curves correspond to different frequencies~(the legend shows $\hbar\omega/k_B T_c$).
The black vertical line indicates the coherence length $\xi_0=2E_F/\pi \Delta k_F$.
For $d$ below this height, the factor $R_1$ plateaus to a constant, while above this height it is decreasing logarithmically with the height.}
    \label{fig:height_dependence}
\end{figure}

\spk{These features are shown in Fig.~\ref{fig:height_dependence}, where we plot 
\begin{equation}~\label{eq:R_1}
    R_{1}=\frac{T_{1,SC}^{-1}}{T_{1,N}^{-1}}
\end{equation}}
where $T_{1,SC}^{-1}$ is the relaxation rate in the superconductor evaluated using Eq.~\eqref{eq:T1} and Eq.~\eqref{eq:current_spectral_density_SC_main}, while $T_{1,N}^{-1}$ is the zero-gap relaxation rate evaluated using Eq.~\eqref{eq:T1} and Eq.~\eqref{eq:normal}.
The figure shows that the relative enhancement $R_1$ is unchanging with height until a scale $d\approx \xi_0$.
Above that height, the factor $R_1$ is suppressed logarithmically with $\ln(d)$ until $d\approx v_F/\omega$.
\spk{For $d\gtrsim \xi_0\hbar\omega E_f/\Delta^2$, the ratio $R_1<1$ indicates a suppression of noise consistent with expectations from a two-fluid model.}

\textit{Discussion.---}
In this work, we have investigated the relaxation of an NV center placed above a 2D superconductor using the $s$-wave BCS theory of superconductivity.
This theory captures the observed coherence peak as the material becomes superconducting. 
This is in contrast to the standard two-fluid model~\cite{Dolgirev_2022,curtis2024probingberezinskiikosterlitzthoulessvortexunbinding}, that predicts a suppression of noise below the superconducting critical temperature.
\spk{The main difference in the two-fluid model is that the normal-fluid particles have a non-singular density of states, while the BCS quasiparticles have the singular density of states.
For an NV close to the sample, $d< \xi_0$, the magnetic noise is strongly sensitive to this singularity.
While for $d\gg \xi_0$, the NV is only sensitive to quasiparticles away from the singularity, and the normal-fluid phenomenology is recovered.
}

Similar to prior works~\cite{zhangFlavorsMagneticNoise2022}, we estimate that current noise, rather than spin noise, is the dominant contribution to the magnetic-field fluctuations probed by the NV.
In the SM, we argue that the magnetic field produced by current noise is large in $(k_F/q)^2$ relative to the field produced by spin noise. 
Thus, spin noise will be relevant only for materials with small Fermi surfaces.
In such cases, we still expect an enhancement in noise as the material becomes superconducting.
This is because both spin and current operators have finite matrix elements for energies at the band gap where the density of states is singular.
In the supplemental~\cite{supplement}, we show any operator that is odd under time-reversal symmetry will be similarly sensitive.
In contrast, operators that are even under time reversal will not be sensitive and not show the subcritical enhancement.
This distinction, previously~\cite{tinkhamIntroductionSuperconductivity1996} labeled ``type I''~(even) and ``type II''~(odd), is why charge-density fluctuations are suppressed, while current fluctuations are enhanced, just below the critical temperature.


Above, we considered a material with thickness $L_z$ small compared to the coherence length.
This allowed us to model the current fluctuations by a two-dimensional BCS model.
Nonetheless, Eq.~\eqref{eq:current_spectral_density_SC_main} holds for any isotropic dispersion and any dimension.
Thus, the requirement $\xi_0>L_z$ could be relaxed if the appropriate three-dimensional dispersion is used.
In this case, the density of states still has a square root singularity and the subcritical enhancement of noise will still occur.

Furthermore, we assumed the mean free path was the longest scale in the system and that we could neglect disorder effects. 
Nonetheless, it is plausible that some of our results hold for an NV height longer than the mean free path, so long as the coherence length is small $\xi_0<l$.
In this limit, noise still originates from quasiparticle states with $\left|\bm{k}\right|\approx k_F\gg d^{-1}$ at which the density of states is singular.
The primary difference is the kinetic constraint is relaxed for $\left|\bm{q}\right|<2\pi/l<2\pi/d$.
This will change the height dependence of the effect, but the subcritical enhancement will likely remain.
This is consistent with results in Ref~\cite{PhysRevB.48.4074}, which investigated the low-frequency conductivity at long-wavelengths $q\rightarrow 0$.
\spk{Lastly, we note that Refs.~\cite{Dolgirev_2022,chatterjeeSinglespinQubitMagnetic2022} uses a phenomenological treatment of strong disorder $l\ll \xi_0$ to justify a two-fluid approach. It's important to highlight that their treatment neglects vertex corrections necessary to capture the coherence peak for $l\gtrsim \xi_0$.}

The lack of the subcritical enhancement in nuclear spin relaxation has been used as a signature of unconventional superconductivity.
In particular, the low-temperature dependence has given evidence to the $d$-wave symmetry of the superconducting order parameter in the cuprates~\cite{Asayama:1996aa}.
Nonetheless, these experiments must account for the modification of the superconducting state when an external magnetic field is applied to polarize the nuclear spins.
In contrast, NVs are outside the sample and can be polarized by optical control.
Thus, future work may find fruitful to consider NV relaxation near unconventional superconductors.

\textit{Acknowledgments-} We greatly appreciate Chunhui Du and Senlei Li for sharing preliminary experimental results and highlighting the enhancement in the relaxation rate below criticality. This work is supported by NSF under Grant No. DMR-2049979.

\bibliography{refs.bib}

\end{document}


\title{Supplemental material for: Superconductivity-enhanced magnetic field noise}
\author{Shane P. Kelly}
\author{Yaroslav Tserkovnyak}

\affiliation{Department of Physics and Astronomy and Mani L. Bhaumik Institute for Theoretical Physics, University of California, Los Angeles, CA 90095}
\maketitle
\onecolumngrid
\vspace{-3em}
\tableofcontents
\section{NV probe of material noise}~\label{sec:matnois}
The mode of NV relaxometry we consider is the $T_1$ noise spectroscopy similar to Refs~\cite{flebusQuantumImpurityRelaxometryMagnetization2018,PhysRevB.98.195433,chatterjeeSinglespinQubitMagnetic2022,Dolgirev_2022,fangGeneralizedModelMagnon2022,zhangFlavorsMagneticNoise2022,PhysRevB.106.115108,zouBellstateGenerationSpin2022,curtis2024probingberezinskiikosterlitzthoulessvortexunbinding,li2024solidstateplatformcooperativequantum,PhysRevB.95.155107}.
In this mode, the NV is prepared, by optical driving, in an excited state, and is allowed to relax to its equilibrium state.
The relaxation dynamics is then probed by optical read out, and the relaxation rate is determined based on the decay of the excited state.
The relaxation is due to the noisy magnetic environment, and in a weak coupling approximation is determined by
\begin{eqnarray}~\label{eq:t1relax}
    T_1^{-1}&=&\left(\frac{g_{\text{nv}}\mu_B}{2\hbar}\right)^2\int_{-\infty}^{\infty}dt e^{i\omega t}\left<\left\{B^{-}\left(\bm{r}_{\text{nv}},t\right),B^{+}\left(\bm{r}_{\text{nv}},0\right)\right\}\right>,
\end{eqnarray}
as introduced in the main text.

In this appendix, we relate the stray magnetic fields produced outside the material to the current and magnetization density inside the material.
Similar calculations have been performed in Refs~\cite{guslienkoMagnetostaticGreenFunctions2011,PhysRevB.95.155107,flebusQuantumImpurityRelaxometryMagnetization2018}, and so we summarize the essential assumptions and aspects of the calculation for completeness.
Our main interest is the magnetic field at position $\bm{r}_{\text{nv}}=\hat{\bm{z}}d+\bm{\rho}$ away from a slab of thickness $L_z$ centered around $z=0$. Here the position is outside the material $d>L_z/2$, and $\bm{\rho}$ is the position in the x-y plane.
For both magnetization and current sources we make use of the Fourier transform:
\begin{eqnarray}
    \frac{1}{\left|\bm{r}_{\text{nv}}-\bm{r}\right|}=\frac{1}{2\pi}\int \frac{d^2\bm{k}}{k} e^{-k\left|d-\hat{\bm{z}}\cdot\bm{r}\right|}e^{i\bm{k}\cdot(\bm{\rho}_{\text{nv}}-\bm{\rho})}.
\end{eqnarray}
\subsection{Demagnetization kernel}
Assuming zero free current, the magnetic field is related to the magnetization via
\begin{eqnarray}
    \bm{\nabla}\cdot\bm{H}=-4\pi\bm{\nabla}\cdot\bm{M},
\end{eqnarray}
which is equivalent to
\begin{eqnarray}
    \bm{B}(\bm{r}_{\text{nv}})=-\int_{\mathcal{R}_3} d^3\bm{r}\bm{\nabla}_{\bm{r}_{\text{nv}}}\left(\bm{\nabla}_{\bm{r}}\frac{1}{\left|\bm{r}_{\text{nv}}-\bm{r}\right|}\right)\cdot\bm{M}(\bm{r})
\end{eqnarray}
for fields outside the material.
Using the above relation we obtain
\begin{eqnarray}
    \bm{B}(\bm{r}_{\text{nv}})=-\frac{1}{2\pi}\int_{\mathcal{R}_2} d^2\bm{k}\int_{\mathcal{R}_3} dzd^2\bm{\rho}\frac{e^{-k\left|d-z\right|}}{k} e^{i\bm{k}\cdot\left(\bm{\rho}_{\text{nv}}-\bm{\rho}\right)}\left(k\hat{\bm{z}}-i\bm{k}\right)\left(-k\hat{\bm{z}}+i\bm{k}\right)\cdot\bm{M}(\bm{r}),
\end{eqnarray}
where we have used $\text{sign}(d-z)=1$ as the NV is above the slab. 
Generally, for arbitrary $d$, the above equation holds under the following replacement:
\begin{eqnarray}
    (k\vechat{z}-i\bm{k})\left(-k\vechat{z}+i\bm{k}\right)\rightarrow\left(k\vechat{z}\,\text{sign}\left(d-z\right)-i\bm{k}\right)\left(k\delta\left(d-z\right)\vechat{z}-k\vechat{z}\,\text{sign}(d-z)+i\bm{k}\right).
\end{eqnarray}
Assuming the magnetization is homogeneous in the $\vechat{z}$ direction: $\bm{M}=\bm{M}_{2}(\bm{\rho})\theta(L_z/2-\left|z\right|)/L_z$, we can perform the integral over $z$:
\begin{eqnarray}
    \bm{B}(\bm{r}_{\text{nv}})=\frac{1}{L_z\pi}\int_{\mathcal{R}_2} d^2\bm{k}\int_{\mathcal{R}_2}d^2\bm{\rho}e^{-kd}\sinh(L_zk/2) e^{i\bm{k}\cdot(\bm{\rho}_{\text{nv}}-\bm{\rho})}\left(\vechat{z}-i\vechat{k}\right)\left(\vechat{z}-i\vechat{k} \right)\cdot\bm{M}_2(\bm{\rho}).
\end{eqnarray}
We therefore write the kernel in momentum space as
\begin{eqnarray}
    \vechat{\bm{G}}_M(\bm{q},d)=4\pi L_z^{-1}e^{-qd}\sinh(L_zq/2) \left(\text{sign}(d)\vechat{z}-i\vechat{q}\right)\left(\text{sign}(d)\vechat{z}-i\vechat{q}\right)
\end{eqnarray}
such that
\begin{eqnarray}
    \bm{B}(\bm{r}_{\text{nv}})=\int_{\mathcal{R}_2} \frac{d^2\bm{q}}{(2\pi)^2}e^{i\bm{q}\cdot\bm{\rho}_{\text{nv}}}\hat{\bm{G}}_M(\bm{q},d)\cdot\bm{M}_2(\bm{q}),
\end{eqnarray}
where
\begin{eqnarray}
    \bm{M}_2(\bm{q})=\int_{\mathcal{R}_2}d^2\bm{\rho}e^{-i\bm{q}\cdot\bm{\rho}}\bm{M}_2(\bm{\rho}).
\end{eqnarray}
The kernel is cutoff at a scale $q>1/d$. Thus, for a thin material, $L_z\ll d$, we have $q<1/d\ll 1/L_z$ and can approximate $L_z q\ll 1$.
This gives the kernel:
\begin{eqnarray}
    \hat{\bm{G}}_M(\bm{q},d)=2\pi qe^{-qd} \left(\text{sign}(d)\vechat{z}-i\vechat{q}\right)\left(\text{sign}(d)\vechat{z}-i\vechat{q}\right)
\end{eqnarray}
\subsection{Oersted kernel}
The magnetic field contributions from the free currents are given by the Oersted kernel
\begin{eqnarray}
    \bm{B}(\bm{r}_{\text{nv}})&=&\frac{1}{c}\int_{\mathcal{R}_3} d^3\bm{r}\frac{\bm{J}(\bm{r})\times(\bm{r}_{\text{nv}}-\bm{r})}{\left|\bm{r}_{\text{nv}}-\bm{r}\right|^3}
    =-\frac{1}{c}\int_{\mathcal{R}_3} d^3\bm{r}\bm{J}(\bm{r})\times\bm{\nabla}_{\bm{r}_{\text{nv}}}\frac{1}{\left|\bm{r}_{\text{nv}}-\bm{r}\right|}.
\end{eqnarray}
Substituting the Fourier transform above and again assuming a homogeneous distribution in the $\vechat{z}$ direction, $\bm{J}(\bm{r})=\bm{J}_2(\bm{r})\theta(\left|z\right|-L_z/2)/L_z$,
we obtain the relation
\begin{eqnarray}
    \bm{B}(\bm{r}_{\text{nv}}))=\int_{\mathcal{R}_2}\frac{d^2\bm{q}}{(2\pi)^2}e^{i\bm{q} \cdot \bm{r}_{\text{nv}}}\bm{G}_J(\bm{q})\times\bm{J}_2(\bm{q})
\end{eqnarray}
with 
\begin{eqnarray}
    \bm{J}_2(\bm{q})=\int_{\mathcal{R}_2}d^2\bm{\rho}e^{-i\bm{q}\cdot\bm{\rho}}\bm{J}_2(\bm{\rho}),
\end{eqnarray}
where the kernel is given as
\begin{eqnarray}
    \bm{G}_J(\bm{q})=\frac{4\pi }{L_zqc}\sinh(L_zq/2)e^{-qd}\left(i\vechat{q}-\text{sign}(d)\vechat{z}\right).
\end{eqnarray}
Again assuming $L_z\ll d$ we obtain 
\begin{eqnarray}
    \bm{G}_J(\bm{q})=\frac{2\pi}{c}e^{-qd}\left(i\vechat{q}-\text{sign}(d)\vechat{z}\right).
\end{eqnarray}
\subsection{NV sensor of material noise}
The $T_1^{-1}$ relaxation rate is related to material noise using Eq.~\eqref{eq:t1relax}, and the magnetic field at the height of the NV as determined above.
The above kernels function then relate the magnetic field correlations to the correlations of the magnetization and current densities.
To simplify these relations, we will assume the NV magnetic dipole moment has orientation perpendicular to the plane, such that $B^{\pm}=B^{x}\pm i B^{y}$.
We will also assume a few properties on the current and magnetization.
First, we assume current fluctuations are dominated by transverse fluctuations $\bm{J}(\bm{q})\cdot\bm{q} \approx 0$ which will be later justified.
Second, we assume the magnetization comes solely from the electron spin and is isotropic.
Under these assumptions the NV relaxation rate due to magnetization noise is given by
\begin{eqnarray}
    T_1^{-1}\propto\int \frac{d^2\bm{q}_1d^2\bm{q}_2}{(2\pi)^4}q_1q_2e^{-(q_1+q_2)d}\int_{-\infty}^{\infty}dt e^{i\Delta t}\left<\left\{M^{a}(\bm{q}_{1},t),M^{a}(\bm{q}_{2},0)\right\}\right>.
\end{eqnarray}

For a finite system that is homogeneous inside the sample, the magnetization fluctuations $\left<\left\{M^{a}(\bm{q}_{1},t),M^{a}(\bm{q}_{2},0)\right\}\right>$ are strongly peaked at $\bm{q}_1=-\bm{q}_2$.
This peak diverges with system size, and in a translation invariant system, correlations of this form~(i.e. $\left<f(\bm{q}_1)f(\bm{q}_2)\right>$) are singular:
\begin{eqnarray}
    \left<f(\bm{q}_1)f(\bm{q}_2)\right>=(2\pi)^2\delta(\bm{q}_1+\bm{q}_2)\int d\bm{r} e^{-i\bm{q}_1\cdot\bm{r}}\left<f(\bm{r})f(0)\right>
\end{eqnarray}
with nonsingular part given by the Fourier transform of the auto correlation function.
The nonsingular part is
\begin{eqnarray}
    \int d\bm{r} e^{-i\bm{q}\cdot\bm{r}}\left<f(\bm{r})f(0)\right>=\lim_{V\rightarrow \infty}\frac{1}{V}\left<f_V(\bm{q})f_V(-\bm{q})\right>=C_{ff}(\bm{q}),
\end{eqnarray}
where $f_V(\bm{q})=\int_V e^{-i\bm{q}\cdot\bm{r}}f(r)$ is the finite volume Fourier transform.

The NV is therefore sensing the power spectral density of the magnetization and current noise.
Explicitly, we find 
\begin{eqnarray}
    T_1^{-1}&=&\left(\frac{g_{\text{nv}}\mu_B}{2\hbar}\right)^2\int d^2\bm{q}e^{-2dq}\left[2q^2C_{\left\{M,M\right\}}(\bm{q},\omega)+ \frac{1}{c^2}C_{\{J_\perp,J_{\perp}\}}(\bm{q},\omega)\right],
\end{eqnarray}
where the power spectral densities of the magnetization and transverse current noise are given as
\begin{eqnarray}
C_{\left\{M,M\right\}}(\bm{q},\omega) &= &  \int dt e^{i\omega t}\lim_{V\rightarrow\infty}\frac{1}{V}\left<\{M^\alpha_V(\bm{q},t)M^\alpha_V(-\bm{q}))\}\right> \\ \nonumber
C_{\{J_\perp,J_{\perp}\}}(\bm{q},\omega)&= &  \int dt e^{i\omega t}\lim_{V\rightarrow\infty}\frac{1}{V}\left<\{J_{\perp,V}(\bm{q},t)J_{\perp,V}(-\bm{q}))\}\right>,
\end{eqnarray}
where $J_\perp(\bm{q})$ is the current density component perpendicular to the wave vector $\bm{q}$.

\subsection{Orientation dependence of NV}
\begin{figure}[h]
    \centering
    \includegraphics[width=2.68in]{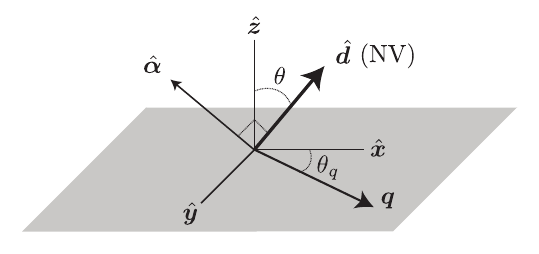}
    \caption{ The projection of $\vechat{d}$ on to the material plane~(shown in gray) is in the $\vechat{x}$ direction. The vector $\bm{q}$ points in plane, and the vectors $\vechat{d}$, $\vechat{\alpha}$, and $\vechat{y}$ form a right-handed basis.}
    \label{fig:angular_dependence}
\end{figure}

We now assume the NV points at an angle $\theta$ relative to normal of the material.
Take the direction of the NV projected onto the material plane as the $\vechat{x}$ direction such that the dipole of the NV is
\begin{eqnarray}
    \vechat{d}=d(\cos\theta \vechat{z} +\sin\theta \vechat{x}) 
\end{eqnarray}
Furthermore, we define $\vechat{y}$ to be perpendicular to the dipole and in the material plane, and define another vector $\vechat{\alpha}$ as
\begin{eqnarray}
    \vechat{\alpha}=\vechat{y}\times \vechat{d}= -\sin\theta \hat{z}+\cos\theta\hat{x},
\end{eqnarray}
to construct a right-handed basis $\vechat{d},\vechat{\alpha},\vechat{y}$. See Fig.~\ref{fig:angular_dependence} for reference.

The relaxation rate due to current fluctuations in directions $\bm{v}_l$ and $\bm{v}_k$ is then
\begin{eqnarray}~\label{eq:t1relax_J}
    T_1^{-1}&=&\left(\frac{g_{\text{nv}}\mu_B}{2\hbar}\right)^2\int \frac{d^2\bm{q}}{(2\pi)^2}\sum_{kl}\left(\bm{n_-}\cdot \bm{G}_J(\bm{q})\times \vechat{v_k}\right)\left(\bm{n_+}\cdot \bm{G}_J(-\bm{q})\times \vechat{v_l}\right)C^{kl}_{\{J,J\}}(\omega,\bm{q}),
\end{eqnarray}
where $\vechat{n_\pm}=\vechat{\alpha}\pm i \bm{y}$

By neglecting contributions from longitudinal-current fluctuations, and only considered transverse-current fluctuations $C_{\{J_\perp,J_\perp\}}=\sum_{kl}(\vechat{q_\perp}\cdot\vechat{v_k})(\vechat{q_\perp}\cdot\vechat{v_l})C^{kl}_{\{J,J\}}(\omega,\bm{q})$, we obtain the relaxation rate as
\begin{eqnarray}~\label{eq:t1relax_angle}
    T_1^{-1}&=&\left(\frac{g_{\text{nv}}\mu_B}{2\hbar}\right)^2\int \frac{q dq d\theta_q}{c^2}e^{-2dq}C_{\{J_\perp,J_\perp\}}(\omega,\bm{q})f(\theta,\theta_q),
\end{eqnarray}
where
\begin{eqnarray}
    f(\theta,\theta_q)=\left(\cos\theta\cos\theta_q+i\left(\sin\theta_q-\sin\theta\right)\right)\left(\cos\theta\cos\theta_q-i\left(\sin\theta_q-\sin\theta\right)\right),
\end{eqnarray}
is the factor that depends on the NV orientation $\theta$, and the angle $\theta_q$ formed by the vector $\bm{q}$ and the $\vechat{x}$ axis (See Fig.~\ref{fig:angular_dependence}).
For an NV pointing in the $\vechat{z}$ direction, $f(\theta,\theta_q)=1$, while for an NV pointing in-plane $f(\theta)=(1-\sin\theta_q)^2$.
When the current noise is isotropic, the integral over $\theta_q$ yields only a geometric factor
\begin{eqnarray}
    F_0(\theta) = \int_0^{2\pi} d\theta_q f(\theta,\theta_q) = \frac{1}{2} \pi  (5-cos (2 \theta)).
\end{eqnarray}

In contrast, if the current fluctuations are anisotropic, depending on $\theta_q$, then the in-plane angle dependence of the NV can be used to resolve that anisotropy.
Suppose now that the material is rotated by an angle $\phi$ around the $\vechat{z}$ axis. The NV relaxation rate is now given as
\begin{eqnarray}~\label{eq:t1relax_angle2}
    T_1^{-1}(d,\omega,\phi)&=&\left(\frac{g_{\text{nv}}\mu_B}{2\hbar}\right)^2\int \frac{q dq d\theta_q}{c^2}e^{-2dq}C_{\{J_\perp,J_\perp\}}(\omega,\bm{q})f(\theta,\theta_q-\phi),
\end{eqnarray}
Decomposing the noise in circular harmonics $C_{\{J_\perp,J_\perp\}}(\omega,q,\theta_q)=\sum_n C_{\{J_\perp,J_\perp\}}(\omega,q,n)e^{i n \theta_q }$, we can compute the geometric factor for each harmonic:
\begin{eqnarray}
    T_1^{-1}(d,\omega,\phi)&=&\left(\frac{g_{\text{nv}}\mu_B}{2\hbar}\right)^2\sum_n F_n(\theta,\phi)\int \frac{q dq }{c^2}e^{-2dq}C_{\{J_\perp,J_\perp\}}(\omega,q,n),
\end{eqnarray}
where $F_n(\theta,\phi)=\int d\theta_q e^{i n\theta_q }f(\theta,\theta_q-\phi)$. Since $f(\theta,\theta_q)$ is real, the geometric factors satisfy $F_n^{\ast}=F_{-n}$.  Since $f$ is a polynomial in trigonometic functions of degree $2$, $F_n(\theta,\phi)=0$ for $\left|n\right|>2$, such that a single NV is only sensitive to the $n\in \left[-2,2\right]$ harmonics. 
The geometric factor for the first and second angular harmonics are
\begin{eqnarray}
   F_1 (\theta, \phi) &=& 2 \pi  \sin \theta  (\sin \phi -i \cos \phi )\\ \nonumber
   F_2 (\theta, \phi) &=& -\frac{1}{2} \pi  e^{2 i \phi } \sin ^2\theta 
\end{eqnarray}
which notably depends on the orientation of the NV.

\section{Ballistic noise in the normal state}\label{apx:bnnm}
In this appendix we review the current and magnetization noise of a noninteracting Fermi gas, and then relate them to the $T_1^{-1}$ relaxation rate of the NV. 
\subsection{Current noise}
For a fixed 2D volume, $V$, the current density at wave vector $\bm{q}$ has the form
\begin{eqnarray}
    \bm{J}(\bm{q})=\int_V d\bm{\rho}e^{-i\bm{q}\cdot\bm{\rho}}\bm{J}(\bm{\rho})=\frac{e\hbar}{m}\sum_{\sigma,\bm{k}} (\bm{k}-\frac{\bm{q}}{2})f^{\dagger}_{\sigma,\bm{k}-\bm{q}}f_{\sigma,\bm{k}},
\end{eqnarray}
and the auto correlation of the magnetic noise is therefore
\begin{eqnarray}
    C_{\{\bm{J},\bm{J}\}}(\bm{q},\omega)=\left(\frac{e\hbar}{m}\right)^2\lim_{V\rightarrow\infty}\frac{1}{V}\sum_{\bm{k},\bm{k}',\sigma,\sigma'}\int dt e^{i\omega t}\left(\bm{k}-\frac{\bm{q}}{2}\right)\left(\bm{k}'+\frac{\bm{q}}{2}\right)\left<\left\{f^{\dagger}_{\sigma,\bm{k}-\bm{q}}(t)f_{\sigma,\bm{k}}(t),f^{\dagger}_{\sigma',\bm{k}'+\bm{q}}f_{\sigma',\bm{k}'}\right\}\right>.
\end{eqnarray}
After applying the commutation rules for fermion operators this expression equates to
\begin{eqnarray}
    C_{\{\bm{J},\bm{J}\}}(\bm{q},\omega)=2\left(\frac{e\hbar}{m}\right)^2\int \frac{d\bm{k}}{2\pi}\delta\left(\omega-E_{\bm{k}-\bm{q}}/\hbar+E_{\bm{k}}/\hbar\right)\left(\bm{k}-\frac{\bm{q}}{2}\right)^2\left( n_{\bm{k}-\bm{q}}\left(1-n_{\bm{k}}\right)+\left(1-n_{\bm{k}-\bm{q}}\right)n_{\bm{k}} \right),
\end{eqnarray}
which scales as $(e\hbar k_F/m_{e})^2=(ev_F)^2$ and with the number of particle-hole pairs near the Fermi surface and with energy difference, $\hbar \omega$ and momenta difference $\hbar \bm{q}$.

The number of such pairs is severely restricted when $\hbar \omega\ll E_F$ and $\hbar \bm{q}\ll \hbar k_F$, which is the case for the NV relaxation process.
In particular, only particles and holes with an energy $\left|E_F-E\right|<k_B T$ near the  Fermi surface contribute, while current fluctuations with $q>1/d$ are suppressed by the kernel relating the magnetic field to the material current.
Thus, we evaluate the constraint by expanding $E_{\bm{k}-\bm{q}}$ around $\bm{k}$: $\omega=\bm{v_g}(\bm{k})\cdot\bm{q}$, where $\bm{v_g}=\grad_{\bm{k}}E_k=\hbar \bm{k}/2m$ is the group velocity.
When $\omega/v_F q$ is small, the constraint enforces $\vechat{k}\cdot\vechat{q}\ll 1$ such that only transverse currents contribute.
For NV frequency in the GHz range, and Fermi velocity $v_F\approx 10^6~\text{m}/\text{s}$, this condition corresponds to $q>10^3~\text{m}^{-1}$.
This condition holds for the wave numbers probed by the NV, $q\approx 1/d\in (10^6, 10^9)~\text{m}^{-1}$.

Under these approximations, we find the current spectral density to be
\begin{eqnarray}
    C^N_{\{J_{\perp},J_{\perp}\}}(\bm{q},\omega)=4(ev_F)^2n_e\frac{k_BT}{E_F}\frac{1}{v_F q},
\end{eqnarray}
where $n_e=k_F^2/2\pi$ is the electron density.  Furthermore, we have assumed $\hbar\omega \ll k_BT$ so that we can approximate $n_{\bm{k}-\bm{q}}\approx n_{\bm{k}}$.

\subsection{Spin noise}
The magnetization carried by the conduction electrons has a form:
\begin{eqnarray}
    \bm{M}(\bm{q})=\int_V d\bm{\rho}e^{-i\bm{q}\cdot\bm{\rho}}\bm{M}(\bm{r})=\mu_B\sum_{\sigma,\sigma',\bm{k}}\bm{\sigma}_{\sigma,\sigma'}f^{\dagger}_{\sigma,\bm{k}-\bm{q}}f_{\sigma',\bm{k}}.
\end{eqnarray}
Thus, the magnetization noise will be similar to the current noise except now the matrix elements are proportional to $\bm{\sigma}$ instead of $\bm{k}-\bm{q}$.
In the limit of small frequencies, $\omega\ll v_F q$, the current matrix elements are fixed $\bm{k}-\bm{q}\approx k_F \vechat{q}_\perp$ perpendicular to the wave vector $\bm{q}$.
Thus, the matrix elements for the current operator are also effectively independent of the momenta of the particle-hole pairs.
The only other difference between the two is that the magnetization noise involves a spin flip, and so there is an overall factor of two relative to the current noise, in which spin is just a degeneracy.
We conclude that the magnetization noise is related to the current noise by a constant factor:
\begin{eqnarray}
    C_{\left\{M,M\right\}}(\bm{q},\omega)=\frac{1}{2}\left(\frac{m\mu_B}{e\hbar k_F}\right)^2C_{\{J_{\perp},J_{\perp}\}}(\bm{q},\omega).
\end{eqnarray}
\subsection{Comparison}
The magnetic field noise produced by the current noise is $N_J(\bm{q},\Delta)=C_{JJ}(\bm{q},\Delta)/c^2$, while that produced by the magnetization noise is $N_M(\bm{q},\Delta)=2q^2C_{MM}(\bm{q},\Delta)$ with a ratio
\begin{eqnarray}
    \frac{N_M(\bm{q},\Delta)}{N_J(\bm{q},\Delta)}=c^2q^2\left(\frac{m\mu_B}{e\hbar k_F}\right)^2=\left(\frac{q}{2k_F}\right)^2.
\end{eqnarray}
Since the magnetic propagator suppresses wave numbers $q>1/d$ and $1/d\ll k_f$, current contribution generally dominates the magnetization contribution.
Spin noise will become relevant if the system has a small Fermi surface~(i.e. $1/k_f \approx 100nm$) and the NV is very close to the sample.
In this case, the small $q$ expansions performed above would be invalid.
Here, we briefly mention a similar comparison was found in Ref.~\cite{zhangFlavorsMagneticNoise2022}, but for an analysis done in the diffusive limit.
In that case, other factors such as the scattering length, and the spin lifetime, can also affect the condition for spin noise to dominate current noise.

\section{BCS Model}

To model the magnetization and current fluctuations, we assume the superconductor can be described by the BCS Hamiltonian
\begin{equation}~\label{eq:BCSham}
    H=\sum_{\bm{k},\sigma}\xi_k f_{\bm{k},\sigma}^{\dagger}f_{\bm{k},\sigma}+\sum_{kl}V_{\bm{k},\bm{l}}f^{\dagger}_{\bm{k},\uparrow}f^{\dagger}_{-\bm{k},\downarrow}f_{-\bm{l},\downarrow}f_{\bm{l},\uparrow}+h.c.\,,
\end{equation}
where $f^{\dagger}_{\bm{k},\sigma}$ is the fermion creation operator with momentum $\bm{k}$ and spin $\sigma$, $\xi_{k}=\hbar^2k^2/2m$ is the dispersion and $V_{\bm{k},\bm{l}}$ is the pairing interaction.
We follow the usual BCS treatment and assume interaction of the form $V_{\bm{k},\bm{l}}=-V\theta(\hbar\omega_D-\xi_{\bm{k}})\theta(\hbar\omega_D-\xi_{\bm{l}})$ where $\omega_D$ is the Debye frequency of the phonons.

Interactions of this form are accurately captured by a mean field approximation in which the $s$-wave superconducting order parameter $\Delta=-\sum_{\bm{l}}V_{\bm{k},\bm{l}}\left<f_{-\bm{l},\downarrow}f_{\bm{l},\uparrow}\right>$, can be determined self consistently.
The mean field Hamiltonian is then diagonalized with quasiparticle energy $E_k=+\sqrt{(\xi_k-E_F)^2+\Delta^2}$, by the canonical transformation 
\begin{eqnarray}~\label{eq:canontrans}
    f_{\bm{k},\uparrow}=u_k \gamma_{\bm{k},\uparrow}+v_k\gamma_{- \bm{k},\downarrow}^{\dagger}\\ \nonumber
    f_{-\bm{k},\downarrow}=u_k \gamma_{-\bm{k},\downarrow}-v_k\gamma_{\bm{k},\uparrow}^{\dagger},
    \label{eq:bdg}
\end{eqnarray}
where $u_k=\cos(\theta_k/2)$, $v_k=\sin(\theta_k/2)$, $E_F$ is the Fermi energy, and $\sin(\theta_k)=\Delta/E_k$ and $\cos(\theta_k)=\xi_k/E_k$.

In the superconducting phase, quasiparticle states only exist for energies above the gap $E>\Delta$ and have a square-root singularity 
\begin{eqnarray}
    D_{SC}(E)=\frac{1}{2}\int \frac{d^2\bm{k}}{(2\pi)^2}\delta(E-E(k))=D_n \frac{E}{\sqrt{E^2-\Delta^2}},
\end{eqnarray}
for relevant $E< \sqrt{E_f^2+\Delta^2}$ and where $D_n$ is the constant $2D$ normal-state density of states $D_n=m/(2\pi\hbar^2)=n_e/E_F=(\hbar v_F 2\pi/k_F)^{-1}$ at the Fermi surface.
Notice, that for $E< \sqrt{E_f^2+\Delta^2}$, the equation $E=E(k)$ has two solutions corresponding to whether $\xi_k-E_F$ is greater or smaller than 0.
The factor of $1/2$ in the definition of $D_{SC}(E)$ is so that only one solution is counted.
This singularity originates from the vanishing group velocity of the superconducting quasiparticles at the Fermi surface $\hbar v_g|_{k_F}=\grad_{\bm{k}} E_k|_{k_F}=0$.

\subsection{Time reversal and the coherence peak}
In this section we show, for an $s$-wave superconductor, that the coherence peak in the noise of an observable requires that the observable is odd under time reversal.
Under time reversal, a general observable
\begin{eqnarray}
    O=\sum_{\bm{k},\bm{k}',\sigma,\sigma'}O_{\bm{k},\bm{k}'}^{\sigma,\sigma'}f^{\dagger}_{\bm{k},\sigma}f_{\bm{k}',\sigma'}
\end{eqnarray}
transforms as
\begin{eqnarray}
    TOT^{-1}= \sum_{\bm{k},\bm{k}'}\sigma\sigma'O_{-\bm{k},-\bm{k}'}^{\bar{\sigma},\bar{\sigma}'*}f^{\dagger}_{\bm{k},\sigma}f_{\bm{k}',\sigma'},
\end{eqnarray}
where $\bar{\sigma}$ indicates the opposite spin from $\sigma\in\{\uparrow,\downarrow\}$, and $T$ is the anti-unitary time-reversal operator.
If the observable is Hermitian, then we have
\begin{eqnarray}
    TOT^{-1}= \sum_{\bm{k},\bm{k}'}\sigma\sigma'O_{-\bm{k}',-\bm{k}}^{\bar{\sigma}',\bar{\sigma}}f^{\dagger}_{\bm{k},\sigma}f_{\bm{k}',\sigma'}.
\end{eqnarray}
Furthermore, if the observable is odd (or even) under time reversal then $TOT^{-1}=\mp O$ and
\begin{eqnarray}~\label{eq:eood}
    O_{\bm{k},\bm{k}'}^{\sigma,\sigma'} = \mp\sigma\sigma'O_{-\bm{k}',-\bm{k}}^{\bar{\sigma}',\bar{\sigma}}.
\end{eqnarray}

In the quasiparticle basis, the observable has the form
\begin{eqnarray}\label{eq:gammaO}
    O=\sum_{\bm{k},\bm{k}',\sigma,\sigma'}\left[O^{\sigma,\sigma'}_{\bm{k},\bm{k}'}u_{\bm{k}}u_{\bm{k}'}-\sigma \sigma'O^{\bar{\sigma}',\bar{\sigma}}_{-\bm{k}',-\bm{k}}v_{\bm{k}}v_{\bm{k}'}\right]\gamma^{\dagger}_{\bm{k},\sigma}\gamma_{\bm{k}',\sigma'},
\end{eqnarray}
where we have neglected pair creation and annihilation $\propto\gamma^{\dagger}\gamma^{\dagger}$ terms.
For energies near the gap, $\left|\bm{k}\right|\approx k_F$, the quasiparticle weights become identical $v_k\approx u_k\approx 1/\sqrt{2}$.
In this region of single-particle Hilbert space, the observable will have significant contributions only if 
\begin{eqnarray}
    O^{\sigma,\sigma'}_{\bm{k},\bm{k}'}-\sigma \sigma'O^{\bar{\sigma}',\bar{\sigma}}_{-\bm{k}',-\bm{k}}\neq 0,
\end{eqnarray}
which is guaranteed by Eq.~\eqref{eq:eood} for observables that are odd under time reversal $TOT^{-1}=-O$, and is impossible for observables that are even under time reversal.
Since the singular density of states occurs for $E\rightarrow \Delta$, only observables which are odd under time reversal will be sensitive to the singularity and show a coherence peak.

\section{Ballistic current noise in a superconductor}~\label{sec:noiseSC}
For a superconductor, the current noise has similar structure to the normal state except for a few key differences.
The first is that, in principle, the current noise in a superconductor has contributions from the breaking and reforming of Cooper pairs.
These contributions only occur when $\hbar \omega> 2\Delta$ which is not the case for the experiment.
In the experiment~\cite{experiment}, the gap is around $10$K, while the NV frequency is around $2$GHz, and so this condition is never met.
The second difference is that the quasi-particles are not directly related to bare electrons.
While the final, most important difference, is that the dispersion has dramatically changed.

Thus, we similarly consider the current spectral density 
\begin{eqnarray}~\label{eq:CJJsuper}
    C_{\{\bm{J},\bm{J}\}}=2\left(\frac{e\hbar}{m}\right)^{2}\int\frac{d^2\bm{k}}{2\pi}\left(\bm{k}-\frac{\bm{q}}{2}\right)^2F_+\left(E_k,E_{k-q}\right)\delta\left(\omega-E_{\bm{k}-\bm{q}}/\hbar+E_{\bm{k}}/\hbar\right)\left[ n_{\bm{k}-\bm{q}}\left(1-n_{\bm{k}}\right)+\left(1-n_{\bm{k}-\bm{q}}\right)n_{\bm{k}}  \right],
\end{eqnarray}
where the coherence factor is given as
\begin{eqnarray}
    F_+(E_k,E_{k-q})=\left(u_{\bm{k}}u_{\bm{k}-\bm{q}}+v_{\bm{k}}v_{\bm{k}-\bm{q}} \right)^2\approx \frac{1}{2}\left(1+\frac{\Delta^2}{E_{\bm{k}}E_{\bm{k}-\bm{q}}}\right),
\end{eqnarray}
and the approximation holds when $\sqrt{E_{\bm{k}}^2-\Delta^2}\sqrt{E_{\bm{k}-\bm{q}}^2-\Delta^2}\ll \Delta^2$.

Instead of immediately expanding the constraint in small $q/k_f$, as done for the normal state, we will first formally integrate $\theta_k$ and make the change of variables from $\left|\bm{k} \right|\rightarrow(E,s)$.
The additional variable $s=\pm1$ is required because the transform from energy coordinates $\xi_{\bm{k}}=E_F+s\sqrt{E^2-\Delta^2}$ is multivalued.
Likewise, the constraint has two solutions for $\sin\theta_{c}$~(4 solutions for $\theta_{c}$) for a fixed $\left|\bm{k}\right|$.
Thus, for a fixed energy $E_{\bm{k}}$, there are at most eight, $c= 1..8$, particle-hole pairs which contribute to the current noise.
For each of these possible contributions, the integral over $\theta_k$ yields 
\begin{eqnarray}
    \int d\theta_k \delta\left(\omega-E_{\bm{k}-\bm{q}}/\hbar+E_{\bm{k}}/\hbar\right)\rightarrow\hbar\left| \partial_{\theta_k}E_{\bm{k}-\bm{q}}\right|^{-1}= D_{SC}(E_{k}+\hbar\omega)\frac{2 \pi \hbar}{k_cq\left|\sin\theta_c\right|},
\end{eqnarray}
where $k_c=\left|\bm{k}\right|(E_k,c)$ is one of the momenta fixed by the energy $E_k$, while $\theta_c$ is one of the solution for $\theta_k-\theta_q$.

Using this change of variables the current spectral density is now
\begin{eqnarray}~\label{eq:current_spectral_density_SC}
    C^{SC}_{\{J_{\perp},J_{\perp}\}}=4\pi(ev_F)^2\int dE D_{SC}(E) D_{SC}(E+\hbar\omega)F_{+}(E,E+\hbar\omega)f_n(E,\hbar\omega,k_BT)\sum_{c}\frac{\hbar\left|\sin\theta_c\right|}{k_cq},
\end{eqnarray}
where $f_n(E,\hbar\omega, k_B T) = \left[ n_{\bm{k}-\bm{q}}\left(1-n_{\bm{k}}\right)+\left(1-n_{\bm{k}-\bm{q}}\right)n_{\bm{k}}  \right]$ contains the dependence on the density of fermions.
This form is similar to the contribution to the spin noise sensed by a nuclear spin~(see below) except for the sum over the eight possible particle-hole pairs, $c$, and the dependence on $\bm{q}$ and $\sin\theta_c$~(also depends on $\bm{q}$).

For $\xi_{\bm{k}}=E_F \left|\bm{k}\right|^2/k_F^2$, the solution to the constraint is given by
\begin{eqnarray}~\label{eq:sol}
    2\tilde{k}_c\tilde{q}\cos\theta_{c}=\tilde{k}_c^2-1+\tilde{q}^2\pm\frac{1}{E_F}\sqrt{(\hbar\omega+E_{\bm{k}})^2-\Delta^2},
\end{eqnarray}
where $\tilde{q}=q/k_F$ and $\tilde{k}_c=k_c/k_F$.
The eight particle-hole pairs come from the two choices for $\tilde{k}_c$, the choice between the $\pm$ solution, and two choices for the sign of $\theta_c$.
Depending on $\bm{q}$ and $\bm{k}$, one or more of these pairs do not have a valid solution for $\theta_c$.
In these limits, the pair does not contribute to the current spectral density.

With these solutions, we find the enhancement of the spectral density of the superconductor's current noise, $R_J(\omega,\bm{q})=C^{SC}_{\{J_{\perp},J_{\perp}\}}/C^N_{\{J_{\perp},J_{\perp}\}}$, relative to the normal-state result is given by
\begin{eqnarray}~\label{eq:enhancementJ}
    R_J(\omega,\bm{q}, T, \Delta)=\frac{1}{4k_BT}\int_\Delta^{\infty}dE \frac{E}{\sqrt{E^2-\Delta^2}}\frac{E+\hbar\omega}{\sqrt{\left(E+\hbar\omega\right)^2-\Delta^2}}F_+(E,E+\hbar\omega )f_n(E) \sum_c \frac{k_F}{k_c}\sin\theta_c,
\end{eqnarray}
while the enhancement of the relaxation rate $R_{1}(\omega,d, T)=T^{-1}_{1,SC}/T^{-1}_{1,N}$ is given by
\begin{eqnarray}
    R_{1}(\omega,d, T, \Delta)=2d\int dq e^{-2dq}R_J(\omega,q, T, \Delta).
\end{eqnarray}

\subsection{The log singularity at low frequency}
For vanishing NV frequency, $\omega\rightarrow 0$, the integrand in Eq.~\eqref{eq:enhancementJ} diverges as $(E-\Delta)^{-1}$ for energies, $E\rightarrow \Delta$ close to the gap.
For finite but small frequencies $\hbar\omega\ll 1$, the integral convergences but is large in $\ln(\hbar\omega/\Delta)$.
This follows by splitting the integral into two parts, one for $E<\Delta/(1-\hbar\omega/\Delta)$, and another for $E>\Delta/(1-\hbar\omega/\Delta)$.
In the first limit, the integrand is approximately constant in energy except for a square root singularity:
\begin{eqnarray}~\label{eq:logint}
    \frac{1}{4 k_BT}F_+(E,E+\hbar\omega )f_n(E) \sum_c\frac{k_F}{k_c}\sin\theta_c\sqrt{\frac{\Delta}{2\hbar\omega}} \int_\Delta^{\frac{\Delta^2}{\Delta-\hbar\omega}}dE \frac{E}{\sqrt{E^2-\Delta^2}}.
\end{eqnarray}
The integral is convergent and not scaling with $\hbar\omega$.
In the second limit, the integrand scales with $E^2/(E^2-\Delta^2)=E^2/(E-\Delta)(E+\Delta)$ and results in the integral scaling logarithmically as $\ln(\Delta/\hbar\omega)$.

\subsection{Height and momentum dependence}
The enhancement of current noise depends on the wave number $\left|\bm{q}\right|$.
In particular, the angle $\theta_c$, and number of particle-hole pairs that solve the kinematic constraint depends on $\bm{q}$~(See main text for a cartoon demonstration).
This dependence also occurs for a normal-state Fermi gas in which two limits arise:
1) when $\left|\bm{q}\right|>2k_F$ and
2) when $\left|\bm{q}\right|<k_F\hbar\omega/E_F$.
In the first limit, the particle-hole pairs cannot be created both on the Fermi surface,
while in the second limit the momentum difference of particle-hole pairs is too small to be separated by an energy $\hbar\omega$.

In the superconductor, the coherence length $\xi_0=2E_F/\pi \Delta k_F$ is a new length scale, above which the phase space for solutions to the kinetic constraint in Eq.~\eqref{eq:sol} is reduced.
Specifically, the phase space is reduced for solutions for which $\bm{k}$ and $\bm{k}-\bm{q}$ are on opposite sides of the Fermi surface.
These type of solutions require
\begin{eqnarray}\label{eq:constraint_ineq}
    E_{\bm{k}}<\frac{1}{2}\sqrt{E_F^2(q/k_F)^2+\Delta^2}
\end{eqnarray}
This constraint to phase space doesn't matter for larger $q$ or for small temperatures, since quasiparticles are suppressed for energies $E_{\bm{k}}\gg k_B T$.
Thus, this constraint is only relevant when $\frac{1}{2}\sqrt{E_F^2(q/k_F)^2+\Delta^2}\approx k_BT\approx \Delta$, or when $q\lesssim \xi_0^{-1}$.

For smaller $q$, Eq.~\ref{eq:constraint_ineq} sets a new upper bound on the noise integral .
The integral still yields a logarithm, but with a new upper bound that is shrinking as $(q\xi_0)^2$.
Thus, we conclude that the scaling of the integral is therefore $\ln((q\xi_0)^2\Delta/\omega)$, which is confirmed by the numerical results for the NV relaxation shown in the main text.

%
%

\subsection{Comparison with nuclear spin relaxation}~\label{sec:nukerelax}
In contrast, nuclear spins relax due to the local spin noise
\begin{eqnarray}
    C_{\{\bm{M},\bm{M}\}}(\bm{\rho}=0, \omega)=\int \frac{d^2 \bm{q}}{(2\pi)^2}C_{\{\bm{M},\bm{M}\}}(\bm{q}, \omega),
\end{eqnarray}
where the important difference is the nuclear spin is sensitive to noise at all length scales.
The spin noise, $C_{\{\bm{M},\bm{M}\}}$, has a similar form to the current density noise in Eq.~\eqref{eq:CJJsuper}.
Upon integral over scattering momenta $\bm{q}$, momentum conservation is relaxed and we find
\begin{eqnarray*}
    \int \frac{d^2 \bm{q}}{(2\pi)^2}C_{\{\bm{M},\bm{M}\}}(\bm{q}, \omega)=8\pi\mu_B^2 \tr\left[\bm{\sigma}\bm{\sigma}\right]\int d E_1 dE_2 D_{SC}(E_1) D_{SC}(E_2)\delta(\hbar\omega-E_1+E_2)F_+(E_1,E_2) f_n\left(E_1,\hbar\omega,k_BT/\Delta\right).
\end{eqnarray*}
Similar to the NV relaxation rate, we can express the ratio between the normal state and superconducting state relaxation as:
\begin{eqnarray}
    \frac{T_{1,SC}^{-1}}{T_{1,N}^{-1}}=\frac{1}{k_BT}\int_\Delta^{\infty}dE\frac{E}{\sqrt{E^2-\Delta^2}}\frac{\left(E+\hbar\omega\right)}{\sqrt{\left(E+\hbar\omega\right)^2-\Delta^2}}\left( 1+\frac{\Delta}{E\left(E+\hbar\omega\right)} \right)f_n\left(E,\hbar\omega,k_BT/\Delta\right).
\end{eqnarray}
Similar to the case of long wavelength magnetic noise, the local spin noise is an integral over a singular density of states, and the integral diverges for $\hbar \omega /\Delta\rightarrow 0$.
The difference is in the lack of height dependence that simplifies the expression.

\bibliography{refs.bib}